\documentclass[prb,showpacs,twocolumn,amsmath,amssymb]{revtex4}   
\usepackage{amsmath}
\usepackage{amsfonts}
\usepackage{amssymb}
\usepackage{graphicx}
\usepackage{epsfig}
\usepackage{bm}

\begin{document}

\title{Ab initio investigation of the melting line of nitrogen at high pressure}

\author{Davide Donadio}
\affiliation{Max Planck Institute for Polymer Research, Ackermannweg 10, 55128 Mainz, Germany.}
\affiliation{Department of Chemistry, University of California Davis} 
\author{Leonardo Spanu}
\affiliation{Department of Chemistry, University of California Davis} 
\author{Ivan Duchemin}
\affiliation{Department of Applied Science,  University of California Davis} 
\author{Francois Gygi}
\affiliation{Department of Computer Science,  University of California Davis} 
\affiliation{Department of Applied Science,  University of California Davis} 
\author{Giulia Galli}
\affiliation{Department of Chemistry, University of California Davis} 
\affiliation{Department of Physics, University of California Davis}

\begin{abstract}
Understanding the behavior of molecular systems under pressure is a fundamental problem in condensed matter physics. In the case of nitrogen, the determination of the phase diagram and in particular of the melting line, are largely open problems. Two independent experiments have reported the presence of a maximum in the nitrogen melting curve, below 90 GPa, however the position and the interpretation of the origin of such maximum differ.
By means of ab initio molecular dynamics simulations based on density functional theory and thermodynamic integration techniques, we have determined the phase diagram of nitrogen in the range between 20 and 100 GPa. We find a maximum in the melting line, 
related to a transformation in the liquid, from molecular N$_2$ to polymeric nitrogen accompanied by an insulator-to-metal transition.
\end{abstract}

\maketitle
The properties of elemental materials under pressure attract 
considerable attention in condensed matter physics,
geophysics and planetary science~\cite{hemley2000,oganov05}.
In particular, nitrogen, with its intricate phase diagram, and its potential applications as an energetic material, has 
been widely studied in recent years; however its phase diagram, including 
its melting line as a function of pressure, is still the subject of 
heated debate.

In its solid form nitrogen remains molecular up to relatively high pressure (P $\sim 100$ GPa) 
and its phase diagram 
exhibits a variety of competing crystalline phases~\cite{Eremets2001}.  
At low P, N$_2$ molecules interact via weak dipolar and quadrupolar forces,
while N atoms are held together by a triple bond, which is 
the strongest chemical bond in nature. The ability of transforming 
the triple bonds of molecular nitrogen 
into single bonds would open the way to storing energy at very high 
density ~\cite{hedm}. This is in principle possible: by pressurizing nitrogen 
to about 110 GPa, non-molecular crystalline and/or amorphous phases are 
formed~\cite{Eremets2001,gregoryanz2001}, as predicted by pioneering theoretical 
works~\cite{mcmahan1985,martin1986}. 
A crystalline covalent form, dubbed ``cubic gauche'' and proposed 
theoretically~\cite{mailhiot1992},  was obtained and fully 
characterized~\cite{cg.eremets} in diamond anvil cell experiments. In 
this insulating phase every N atom forms three single covalent 
bonds with its neighbors arranged in a cubic lattice. Further covalent 
crystalline forms have been predicted to occur at even higher 
P~\cite{ma2009, pickard2009}, but are yet to be found in 
experiments. 

By analogy with high pressure solid phases, the existence of non-molecular, 
liquid nitrogen was suggested as well~\cite{ross1987}, and 
very recently a first order liquid-liquid phase transition has been proposed~\cite{bonev}, between a low density liquid
molecular phase (LDL) and a high density liquid polymeric phase (HDL), located 
between 2000 and 6000 K at $\sim 80$ GPa. Such structural 
transformation is accompanied by metallization of fluid nitrogen, as
observed in shock reverberation experiments~\cite{chau2003}.
Though uncommon in elemental 
liquids, a first-order liquid-liquid (LL) phase transition, from a 
molecular to an atomic phase, has been observed in 
phosphorus~\cite{katayama2000}, which is isovalent to nitrogen. 
Support in favor of a LL phase transition in nitrogen comes from the 
observation of a maximum in the melting curve~\cite{boehler, goncharov}, 
whose  presence may be an indication of a change in the liquid properties 
~\cite{bonevH2,sodium,katayama2000}. 
A negative slope of the melting 
line may also be associated with the presence of open crystalline structures,
as e.g. in carbon and water, or with changes in the  electronic structure of the 
system, for example a metal-insulator transition~\cite{sodium} or promotions 
of valence electrons to electronic orbitals higher in energies than those 
occupied at low P, as found, e.g. in Cesium.

The position and the character of the maximum in the melting 
curve of nitrogen are still matter of debate~\cite{boehler, goncharov,comment1,comment2}
According to Ref.~\cite{boehler} the maximum is 
sharp and located at 50 GPa, and may possibly be the signature of a 
triple point associated to a first order LL phase transition.
Goncharov {\it et al.}~\cite{goncharov,comment1} measured instead a 
slight change in the melting line slope near $70$ GPa. In addition, by 
performing in situ Raman scattering, they found no evidence of a LL phase 
transition, and related the maximum in the melting curve to polymorphic 
transitions between crystalline molecular phases.

In this Letter we report the theoretical melting line of nitrogen between 
20 and 100 GPa as obtained from first principle molecular dynamics 
(MD) simulations. We predict that the melting temperature reaches a maximum 
between 80 and 90 GPa, in correspondence to a transition in the liquid phase 
involving both a structural
modification from a molecular to a polymeric fluid, and a semiconductor to metal transition. 
We show that close to the maximum,
the liquid polymerizes and becomes denser than the 
corresponding molecular solid, thus giving rise to a negative slope in 
the PT melting curve.

Calculations of melting lines can be obtained either by the two-phase 
simulation method~\cite{bonevH2,correa2005,water} or by thermodynamic 
integration (TI). 
The two approaches are in principle equivalent~\cite{ourjcp}, but the 
two-phase method may require larger simulation cells and longer runs to 
achieve accuracy comparable to TI.  We therefore 
determined the melting temperature of nitrogen at several different 
P by computing free energy differences between liquid and 
crystalline phases by TI, in a manner similar to previous studiesi of C and Si melting lines~\cite{sugino1995,wangscandolo}.
Our computational framework relies upon Born-Oppenheimer MD
simulations~\cite{qbox,gygi2009}, where the electronic structure is 
solved within density functional theory (DFT). We used a generalized 
gradient approximation, PBE ~\cite{pbe}, for the exchange and correlation 
functional, norm conserving pseudopotentials and a plane-wave expansion 
of the electronic orbitals with a kinetic-energy cut-off of 60 Ry. We 
simulated nitrogen in supercells containing 128 atoms with $\Gamma$-point 
sampling of the supercell Brillouin zone. The MD equations of motion are 
integrated with a time-step of 20 a.u., the temperature is controlled by 
stochastic velocity rescaling~\cite{bussi} and the pressure is kept 
constant by first-order cell dynamics.
This scheme yields good agreement with the P(V) curve of liquid N at 2000 K 
in Ref.~\cite{bonev}.

The use of TI requires the availability of a potential to describe a 
reference system, for which we have chosen a classical force-field that provides a good description 
of molecular nitrogen at low P.\cite{potential}\footnote{This potential is made of an 
intra-molecular Morse term and an intermolecular Lennard-Jones (LJ) 
term. We have rescaled the LJ parameters, so as to 
bring the melting temperature of the reference system closer the 
experimental one, as a function of P.-end move}.
Our TI protocol consists of three steps: 
(i) We compute the melting temperature of the reference system 
(T$^{ref}_m$) at a given pressure P by a two-phase simulation, using  a supercell 
with 3400 molecules and a simulation time of 100 ps. We note that the 
melting line of the classical system is monotonic as a function of P and 
does not exhibit any maximum. 
(ii) The (Gibbs) free energy differences ($\Delta G$) of the reference and the 
DFT/PBE systems are computed both for the solid and the liquid phase by 
adiabatic switch~\cite{watanabe1990} (AS) in 2 ps runs. 
The convergence of $\Delta G$ as a function of switching time has been tested by performing
longer AS runs at 40 GPa and error bars on $\Delta G$ are obtained as the standard deviation
of four independent runs. 
Longer AS runs, up to 10 ps have been performed to obtain an 
accurate evaluation of the free energy of the liquid at 90 GPa, given the 
inability of the classical reference model to describe the dissociated liquid.
The free energy differences between solid and liquid at different volumes 
need to be corrected because of the finite plane wave cutoff used in our 
simulations~\cite{sugino1995}. A correction term is computed by 
performing single point calculations for selected MD snapshots with a plane 
wave cutoff of 160 Ry, that is sufficient to yield well converged values 
of P at all volumes considered here.
 (iii) The melting temperature (T$_m$) of nitrogen at the PBE level is 
finally computed by reversible scaling~\cite{dekoning1999,dekoning2001}
\footnote{The force rescaling factor is varied linearly between 1 and 0.7, over 4 ps long MD simulations.}: 
T$_m$ is located at the intersection of the Gibbs free energy curves 
(${\cal G}(T)$) of the liquid and the solid phases, with an initial 
offset determined by $\Delta{\cal G}$ at T$^{ref}_m$, obtained in step (ii). 
In addition, we have computed the slope of the melting curve at 65, 80.5 
and 90 GPa using the Clausius-Clapeyron equation~\cite{kofke1993}, 
evaluating the differences in density and enthalpy over 5 ps long independent {\sl ab initio} MD 
runs, carried out in the canonical ensemble at constant pressure (NPT). 

\begin{figure}
[ptb]
\begin{center}
\includegraphics[width=2.9in]{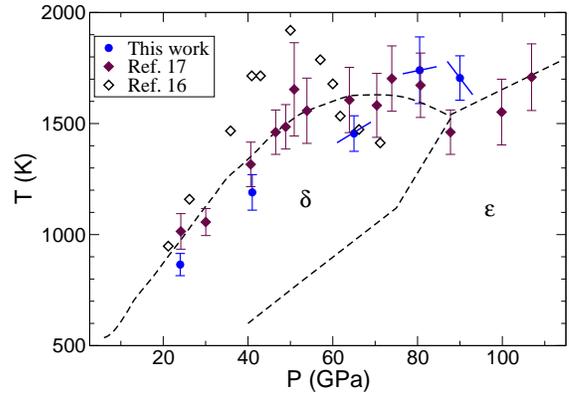}
\caption{(color online) Proposed phase diagram of nitrogen. The computed points on the melting line
are indicated with blue circles and the slopes obtained from the 
Clausius-Clapeyron equation with blue straight lines. The experimental points from 
Goncharov {\it et al.}~\cite{goncharov} are shown as solid diamonds and 
those measured by Mukherjee and  Boehlr~\cite{boehler} as empty diamonds. 
}
\label{Fig:PT}
\end{center}
\end{figure}
Using the procedure discussed above, we have computed the melting 
temperature at five different values of P: the results are 
reported in Fig.~\ref{Fig:PT} and compared with the experiments from 
Refs.~\cite{goncharov,boehler}. Our results agree 
within one error bar with the melting points measured by Goncharov {\it 
et al.}~\cite{goncharov}, while they are not compatible with the presence 
of a cusp at 50 GPa, as in Ref.~\cite{boehler}. Our computed 
melting curve displays a maximum between 80.5 and 90 GPa; the presence of 
a maximum is further supported by the opposite signs of the slope of the 
curve, computed at 80.5 and 90 GPa. The position of the maximum in the 
melting line is shifted toward higher pressure with respect to the 
measurements of Ref.~\cite{goncharov} by about 10 GPa.

Our simulations show that the thermodynamically stable crystalline phase between 80 and 90 GPa is molecular,
even at T close to the melting point, 
in agreement with the experimental observations in ~\cite{Eremets2001, goncharov}. 
Yet, at variance with Ref.~\cite{goncharov}, we could not locate a polymorphic phase transition
between the $\delta$ and the $\varepsilon$ phase  within this pressure range.
By direct {\sl ab initio} MD simulations in the NPT ensemble we observe that at 1500 K and 90 GPa phase $\varepsilon$ transform rapidly into 
$\delta$ and we estimate the $\delta / \varepsilon$ phase boundary at P above 120 GPa at 1500 K.
Experiments confirm that solid nitrogen is molecular in the (P,T) range of interest (80-90 GPa, $\sim 1750$K),
however the presence of several competing metastable structures and strong hysteresis effects
make the determination of the stable crystalline phase uncertain~\cite{goncharovprivcomm}. 
Nevertheless the density differences between 
the various molecular solids observed at this P are small, compared to the density difference between the liquid and the 
solid phase; thus uncertainties in the relative stability of the molecular polymorphs may only result in small quantitative variations of the melting line predicted by our simulations.

Therefore we conclude that the presence of a maximum in the melting line 
stems from structural and electronic transformations 
occurring in the liquid, rather than in the solid phase. As first observed in Ref.~\cite{bonev}, and confirmed by our simulations, liquid nitrogen undergoes a transition from a 
molecular to a polymeric phase, which at 2000 K occurs at $\sim$88 GPa. 
The analysis of our liquid sample at 90 GPa shows 
the coexistence of molecular N$_2$ and chains of N atoms where triple 
bonds give way to longer single bonds, whose signature appears as a 
second peak in the radial distribution function (not shown) at $\sim$1.3 \AA , while
the triple bond yields a peak at 1.1 \AA .
The distribution of the bond 
angles around the tetrahedral value (109.3) and the observed tendency to 
form pentagonal rings are signatures of sp$^3$ 
hybridization of the atomic orbitals, analogous to the one observed in 
covalent crystalline phases~\cite{cg.eremets, ma2009} of nitrogen. The 
formation of covalent chains causes a drop in the volume of the liquid, 
which becomes denser than its crystalline counterpart. At 90 GPa and at 
the predicted melting temperature of 1705 K, liquid N is 1$\%$ denser 
than the solid. Such density difference results in a negative slope of 
the melting line. 

\begin{figure}
[ptb]
\begin{center}
\includegraphics[width=2.9in]{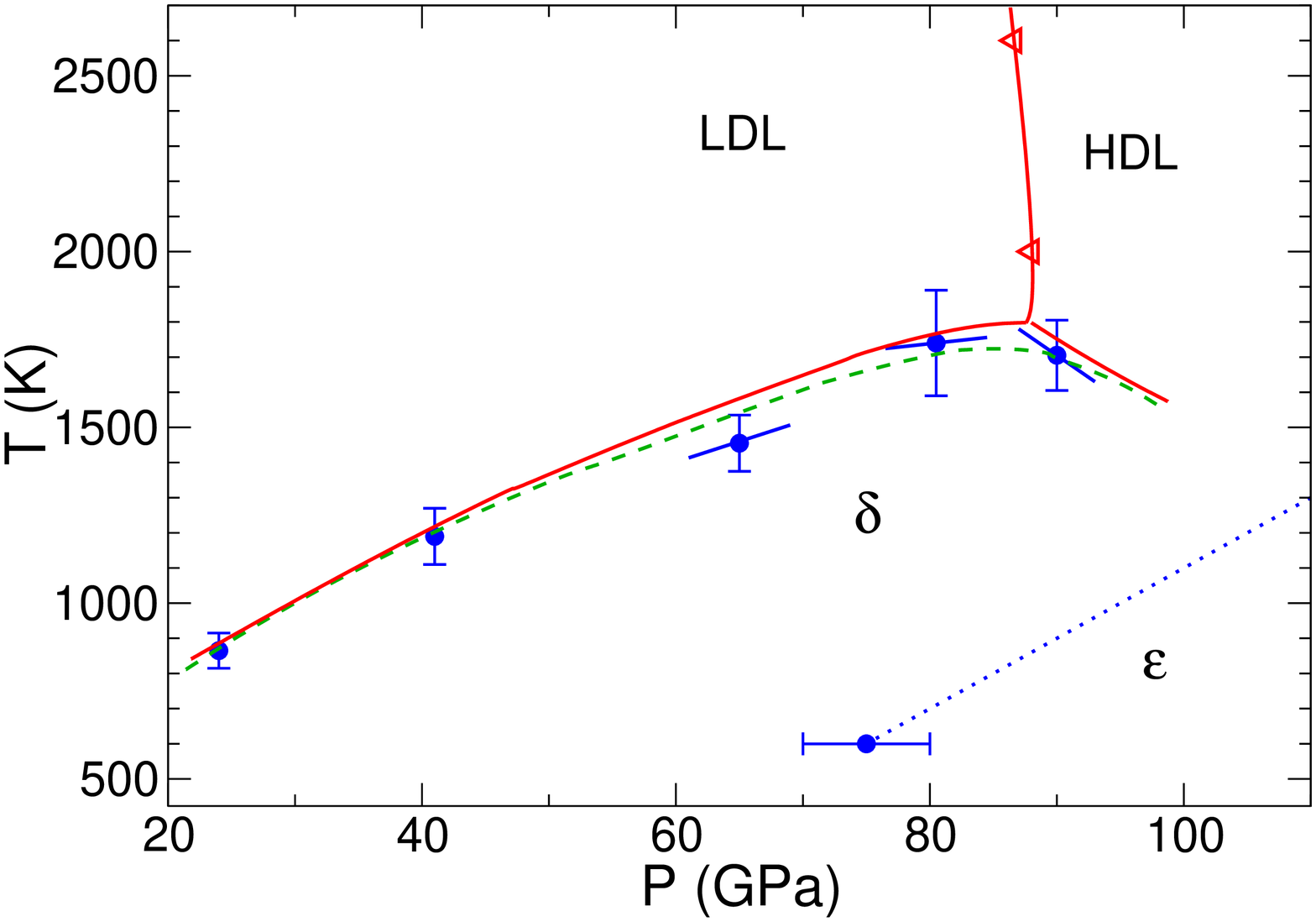}
\caption{(color online). 
Computed melting line of nitrogen (see Fig.1) and liquid-liquid phase boundary. 
The red solid line (guide to the eye) indicates the presence of a triple point, 
occurring in the case of a first-order LL phase transition.
Instead, the green dotted line (guide to the eye) does not 
show any triple point, corresponding to a second order LDL-HDL transition. 
}
\label{phase-diagram}
\end{center}
\end{figure}

If the molecular to polymeric LL transition was first order, as suggested in Ref.~\cite{bonev}, 
then the maximum in the melting curve predicted by our simulations would  
coincide with a triple point, and it would be a cusp (i.e. the melting line would have discontinuous derivatives at the maximum), as 
Ref.~\cite{boehler}. However the location of the maximum found here is different from the one 
found in Ref.~\cite{boehler} (at about 50 GPa). If the LL transition was instead second order, then the derivative of the melting line would be a continuous function, similar to what observed, for example, in liquid carbon~\cite{correa2005,wangscandolo}. The 
two possible scenarios are illustrated in Fig.~\ref{phase-diagram}.
We note that the characterization of the LL phase transition as first-order~\cite{bonev} 
is plausible but not definitive: the use of small 
periodic simulation cells may make a second-order phase transition 
resemble a first-order one, and a statistically 
significant size scaling analysis was not performed~\cite{bonev}. Unfortunately such analysis would likely require cells with at least 500 molecules and, most importantly, simulation times of the order of several ns, that are outside the reach of current ab initio simulation techniques.

\begin{figure}
[ptb]
\begin{center}
\includegraphics[height=2.2in,width=2.5in]{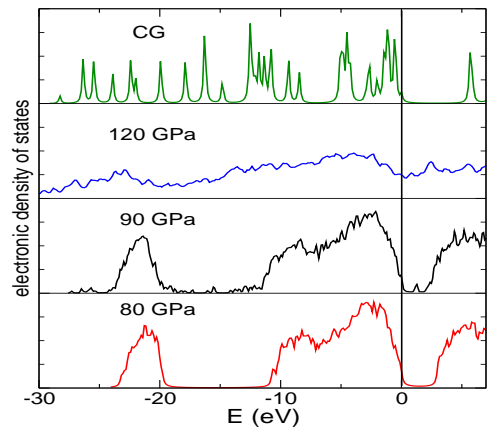} 
\caption{(color online) Computed electronic density of states for the liquid phase at $80$, $90$ 
and $120$ GPa and for the Cubic Gauche (CG) crystalline phase.}
\label{edos}
\end{center}
\end{figure}

\begin{figure}[ptb]
\begin{center}
\includegraphics[height=2.2in,width=2.4in]{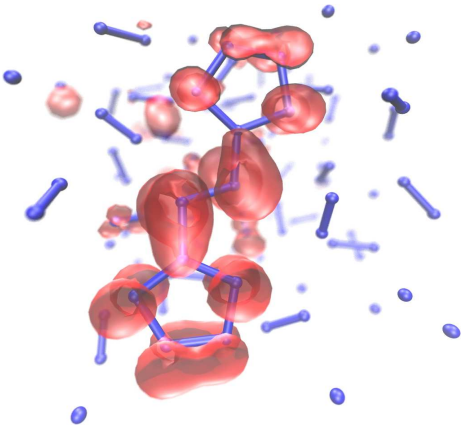}
\caption{(color online) Isosurface of the square modulus of the wavefunction of a single particle state with energy inside the electronic gap of liquid nitrogen at 90 GPa.
 }
\label{density}
\end{center}
\end{figure}

As a consequence of the structural changes occurring upon compression, the electronic structure of the 
liquid undergoes major modifications (Fig.~\ref{edos}). Up to 80 GPa 
liquid nitrogen is an insulator with a DFT/PBE gap of 3.1 eV. 
At 90 
GPa, the formation of chains and pentagonal rings give rise to the appearance of defect-like states in the middle of 
the electronic gap. These states have anti-bonding $\pi^*$ character and are 
delocalized over the polymeric chains (Fig.~\ref{density}). 
As P is further increased, the liquid loses its molecular 
character, an increasing number of polymeric chains are formed, and eventually metallization 
occurs. The electronic density of states of liquid nitrogen at 120 GPa 
shows indeed no gap. 
The observed metallization is consistent with an increase in electrical 
conductivity measured in shock reverberation experiments~\cite{chau2003}.
The density of states of the CG phase at 120 GPa is shown in 
the upper panel of Fig.~\ref{edos}: it is remarkable that the stable 
crystalline covalent polymorphs of nitrogen are semiconducting 
(or insulating) up to a pressure as high as 240 GPa~\cite{Eremets2001}. 
However a chain-like metallic crystalline polymorph was predicted to have 
an enthalpy close to that of the CG phase~\cite{Mattson2004}. 

In summary, by means of {\sl ab initio} MD simulations and thermodynamic integration 
we have determined the melting line (T$_m$ (P)) of nitrogen up to 90 GPa. We have found that T$_m$ (P) exhibits a maximum between 80 and 90 GPa which is related to a structural transformation in the liquid, from a molecular to a polymeric phase. This transformation is accompanied by an insulator to metal transition. Our computed melting temperatures are in fair agreement with those determined in recent diamond anvil cell 
experiments~\cite{goncharov}, and not compatible with the data of Ref.~\cite{boehler}, where melting was established by visual inspection. If the transformation observed in the liquid corresponds to a first order phase transition, as suggested in Ref.~\cite{bonev}, then the maximum found here will coincide with a triple point and thus a cusp in the melting line, as proposed by Ref.~\cite{boehler}. However, the shift in the transition of the electronic properties of the liquid, which undergoes metallization at high pressure, with respect to its structural transition is an indication against a first order transition in the liquid phase. Work is in progress to compute spectroscopic observables capable of unequivocally identifying different liquid phases.

We thank S.~A. Bonev and T.~Ogitsu for useful discussions. This work was supported dy DOE/Scidac through grant No. DE-FG0206ER46262 and by NSF under OCI PetaApps grant 0749217.

\end{document}